\documentclass[12pt]{article}
\usepackage[T1]{fontenc}
\usepackage[utf8]{inputenc}
\usepackage[dvipsnames]{xcolor}
\usepackage{colortbl}
\usepackage{babel}
\usepackage{array}
\usepackage{booktabs}
\usepackage{mathtools}
\usepackage{url}
\usepackage{multirow}
\usepackage{amsmath}
\usepackage{amsthm}
\usepackage{amssymb}
\usepackage{graphicx}
\usepackage{geometry}
\usepackage{setspace}
\usepackage[authoryear,round]{natbib}
\usepackage[pdfusetitle,
 bookmarks=true,bookmarksnumbered=false,bookmarksopen=false,
 breaklinks=true,pdfborder={0 0 0},pdfborderstyle={},backref=false,colorlinks=true]
 {hyperref}
\hypersetup{
 citecolor=blue}

\makeatletter

\providecommand{\tabularnewline}{\\}

\usepackage{titling}
\usepackage[compact]{titlesec}
\date{}

\makeatother

\begin{document}
\title{\textbf{Accelerating Birkhoff Projection for Manifold-Constrained
Hyper-Connections}}
\author{Chenrui Wang\\
{\normalsize School of Statistics}\\
{\normalsize Renmin University of China}\\
{\normalsize\texttt{iamwangchenrui@gmail.com}}\\\\
Yixuan Qiu\\
{\normalsize School of Statistics and Data Science \& Institute of
Big Data Research}\\
{\normalsize Shanghai University of Finance and Economics}\\
{\normalsize\texttt{qiuyixuan@sufe.edu.cn}}}
\maketitle

\begin{abstract}
Manifold-constrained hyper-connections (mHCs) have recently been proposed
as a principled extension of hyper-connections, where the residual
mixing matrices are constrained to be doubly stochastic via projection
onto the Birkhoff polytope. In practical mHC implementations, this
constraint is enforced by Sinkhorn--Knopp iterations, and the backward
pass relies on unrolling the iterative solver. This design introduces
substantial computation and memory overhead, and may also yield inaccurate
projections when the algorithm converges slowly on challenging inputs,
undermining the intended norm-control and stability guarantees of
mHCs.

In this work, we focus on the practically important $4\times 4$
Birkhoff projection setting and develop an end-to-end acceleration framework.
By leveraging the dual formulation, we reduce the problem to a three-dimensional
unconstrained convex problem and solve it with Newton's method, achieving
fast convergence and high accuracy. For the backward pass, we replace
the unrolled differentiation with implicit differentiation, yielding
exact gradients without storing intermediate states. To exploit massive
parallelism, we design a warp-level CUDA kernel that uses only register-level
primitives, avoiding global and shared memory I/O.

Extensive experiments against representative open-source baselines
demonstrate that the proposed solver yields substantially more reliable
doubly stochastic projections---especially when the input magnitude
is large---and achieves significant end-to-end speedups (including
the backward pass), reaching over $20\times$ acceleration at large
batch sizes while maintaining orders of magnitude smaller marginal
errors.
\end{abstract}

\section{Introduction}

In recent years, the evolution of deep neural network architectures
has witnessed a renewed interest in architecture design beyond the
conventional residual connection paradigm. The introduction of hyper-connections
(HC, \citealp{zhu2025hyper}) marked a significant step forward by
expanding the residual stream width and enabling learnable, multi-path
information flow across layers. While HC demonstrated substantial
performance gains in large-scale language model pre-training, it also
introduced critical challenges: the unconstrained nature of its residual
mappings compromises the identity mapping property, leading to training
instability and limited scalability.

To address these issues, manifold-constrained hyper-connections (mHC,
\citealp{xie2025mhc}) was recently proposed as a principled extension
that projects the residual mappings onto the Birkhoff polytope---the
set of doubly stochastic matrices. By enforcing row and column sum
constraints, mHC restores the signal conservation property that is
essential for stable training, while preserving the expressive power
of multi-stream architectures. 

In practice, mHC employs the Sinkhorn--Knopp algorithm \citep{sinkhorn1964relationship,sinkhorn1967concerning}
to perform this projection iteratively. However, for the typical expansion
rates used in mHC (e.g., $n=4$ or 8), the Sinkhorn iterations introduce
non-negligible computational overhead, especially when executed for
every token across millions of training steps. Moreover, the backward
pass in automatic differentiation frameworks typically requires unrolling
the entire iterative process, leading to significant memory and computational
costs. 

Another potential concern for the current implementation of mHC is
that the Sinkhorn--Knopp algorithm may converge slowly in challenging
problems, as existing works have pointed out \citep{yin2025wasserstein,wu2025pins,chhaibi2025faster}.
Therefore, using a fixed and small number of Sinkhorn--Knopp iterations
(e.g., 20 iterations as suggested by \citealp{xie2025mhc}) may be
insufficient.
Inadequate convergence can have several detrimental effects
on the output residual mapping: it is not guaranteed to be a doubly
stochastic matrix, and its operator norm is uncontrolled. Recall that
one of the main motivations of mHC is to stabilize the matrix norms
in HC, so designing algorithms that achieve a high accuracy of Birkhoff
projection while requiring minimal computational cost is a crucial
part of reliable and efficient mHC implementations.

In this work, we focus on accelerating the projection operator at
the heart of mHC, with an emphasis on the practically important case
$n=4$. Our key observation is that the Birkhoff projection problem
considered by mHC is mathematically equivalent to solving an entropic-regularized
optimal transport (OT) problem \citep{cuturi2013sinkhorn}, which
is extensively studied in the literature. Importantly, the $4\times4$
Birkhoff projection induced by entropic-regularized OT admits a three-dimensional
dual formulation, enabling a second-order method with fast convergence.
Building on this structure, we develop an end-to-end acceleration
framework that improves both the forward projection and backward differentiation.
Specifically, we make the following contributions:
\begin{enumerate}
\item \textbf{Forward pass via Newton's method}: We reformulate the dual
of the entropic OT problem as an unconstrained convex optimization
in $\mathbb{R}^{3}$, and derive closed-form expressions for the gradient
and Hessian. This enables the use of Newton's method, which converges
quadratically and typically requires far fewer iterations than Sinkhorn.
\item \textbf{Backward pass via implicit differentiation}: Instead of backpropagating
through the iterative solver, we derive an analytical expression for
the derivative of the projection using the implicit function theorem.
This allows us to compute gradients exactly and efficiently, without
storing intermediate iterates.
\item \textbf{GPU-efficient implementation}: We design a warp-level CUDA
kernel that processes two $4\times4$ matrices simultaneously using
only register-level primitives. The implementation avoids shared memory
and global memory I/O, achieving high throughput with minimal overhead.
\end{enumerate}

\section{Background}

\subsection{Manifold-Constrained Hyper-Connections}

HCs extend the classical residual connection \citep{he2016deep} by
expanding the residual stream from a single vector to multiple parallel
streams. For the $l$-th layer, let $\mathbf{x}_{l}\in\mathbb{R}^{n\times C}$
denote the input hidden matrix, where $C$ is the input dimension
and $n$ is the expansion rate. HC introduces three learnable linear
mappings: $\mathcal{H}^{\mathrm{pre}}_{l}\in\mathbb{R}^{1\times n}$
that aggregates the streams into a single input for the layer function
$\mathcal{F}_{l}$, $\mathcal{H}^{\mathrm{post}}_{l}\in\mathbb{R}^{1\times n}$
that maps the layer output back to the streams, and $\mathcal{H}^{\mathrm{res}}_{l}\in\mathbb{R}^{n\times n}$
that mixes the streams after the residual addition. The forward pass
is given by:
\[
\mathbf{x}_{l+1}=\mathcal{H}^{\mathrm{res}}_{l}\mathbf{x}_{l}+\left(\mathcal{H}^{\mathrm{post}}_{l}\right)^{T}\mathcal{F}_{l}\left(\mathcal{H}^{\mathrm{pre}}_{l}\mathbf{x}_{l}\right).
\]

While HC significantly improves model performance by enabling richer
cross-layer information flow, it also introduces a critical drawback:
the repeated application of unconstrained matrices $\mathcal{H}^{\mathrm{res}}_{l}$
across layers leads to the composite mapping $\prod^{L-1}_{i=l}\mathcal{H}^{\mathrm{res}}_{L-i}$,
whose norm may explode or vanish, violating the identity mapping principle
that is essential for stable gradient propagation \citep{he2016deep}.
As a result, HC suffers from training instability and limited scalability
in large-scale settings \citep{xie2025mhc}.

To remedy this, \citet{xie2025mhc} proposes mHC that enforces the
residual mixing matrices $\mathcal{H}^{\mathrm{res}}_{l}$ to be doubly
stochastic, \emph{i.e.}, they belong to the Birkhoff polytope:
\[
\mathcal{B}_{n}=\{P\in\mathbb{R}^{n\times n}:P\mathbf{1}_{n}=\mathbf{1}_{n},P^{T}\mathbf{1}_{n}=\mathbf{1}_{n},P\ge0\},
\]
where the inequality $P\ge0$ means that every entry of $P$ is nonnegative.
Doubly stochastic matrices have several desirable properties: their
spectral norms are bounded by one, they are closed under matrix multiplication,
and they act as convex combinations of the input streams, thereby
preserving the average signal magnitude. Consequently, the composite
mapping across multiple layers remains well-behaved, restoring the
identity mapping property and ensuring stable training.

\subsection{The Sinkhorn--Knopp Algorithm}

One of the key innovations of mHC compared to HC is to project an
unconstrained residual mapping $\exp(\tilde{\mathcal{H}}^{\mathrm{res}})$
onto the Birkhoff polytope $\mathcal{B}_{n}$ via the Sinkhorn--Knopp
algorithm, where the exponential function $\exp(\cdot)$ applies to
each element of its matrix argument. Sinkhorn's theorem \citep{sinkhorn1964relationship}
states that any $n\times n$ matrix $A$ with strictly positive entries
can be transformed into a doubly stochastic matrix $T$ by proper
row and column scalings, \emph{i.e.}, there exist diagonal matrices
$D_{1}$ and $D_{2}$ with strictly positive diagonal elements such
that $T=D_{1}AD_{2}$. Moreover, $D_{1}$ and $D_{2}$ are unique
up to a scalar multiplier.

The matrix $T$ can be viewed as the projection of $A$ onto the Birkhoff
polytope $\mathcal{B}_{n}$ in a specific geometry. In fact, define
the generalized Kullback--Leibler (KL) divergence between two matrices
$A,B\in\mathbb{R}^{n\times m}_{+}$ with positive entries as
\[
\mathrm{KL}(B\Vert A)=\sum_{i,j}\left\{ B_{ij}\log\left(\frac{B_{ij}}{A_{ij}}\right)-B_{ij}+A_{ij}\right\} ,
\]
and then we can show that $T=D_{1}AD_{2}$ minimizes the KL divergence
between $A\in\mathbb{R}^{n\times n}_{+}$ and a doubly stochastic
matrix:
\begin{equation}
T=\mathrm{proj}^{\mathrm{KL}}_{\mathcal{B}_{n}}(A)\coloneqq\underset{P\in\mathcal{B}_{n}}{\arg\min}\,\mathrm{KL}(P\Vert A).\label{eq:birkhoff_projection}
\end{equation}

One simple iterative method to find such $D_{1}$ and $D_{2}$ matrices
is the Sinkhorn--Knopp algorithm \citep{sinkhorn1967concerning},
which alternately rescales the rows and columns of $A$ to sum to 1.
For brevity of notation, let $\oslash$ denote the elementwise division
between vectors. Given an initial value $v^{(0)}\in\mathbb{R}^{n}$,
the Sinkhorn--Knopp algorithm can be described by the following procedure:
\begin{equation}
u^{(k+1)}=\mathbf{1}_{n}\oslash(Av^{(k)}),\quad v^{(k+1)}=\mathbf{1}_{n}\oslash(A^{T}u^{(k+1)}),\quad k=0,1,\ldots.\label{eq:sinkhorn_knopp}
\end{equation}
As $k\rightarrow\infty$, $\mathbf{diag}(u^{(k)})$ and $\mathbf{diag}(v^{(k)})$
will converge to $D_{1}$ and $D_{2}$, respectively.

\subsection{Entropic-Regularized Optimal Transport}

\label{subsec:entropic_regularized_ot}

The Birkhoff projection problem (\ref{eq:birkhoff_projection}) and
the Sinkhorn--Knopp algorithm (\ref{eq:sinkhorn_knopp}) have a strong
connection with the entropic-regularized OT problem, which can be
characterized by the following optimization problem:
\begin{equation}
\min_{T\in\Pi(a,b)}\langle T,M\rangle-\eta\cdot h(T),\label{eq:entropic}
\end{equation}
where $M\in\mathbb{R}^{n\times m}$ is a given cost matrix, $a$ and
$b$ are two probability vectors satisfying $a>0$, $b>0$, and $\sum^{n}_{i=1}a_{i}=\sum^{m}_{j=1}b_{j}=1$,
$\eta>0$ is a regularization parameter, $h(T)=-\sum_{i,j}T_{ij}(\log(T_{ij})-1)$
is the entropy term, and
\[
\Pi(a,b)=\{P\in\mathbb{R}^{n\times m}:P\mathbf{1}_{m}=a,P^{T}\mathbf{1}_{n}=b,P\ge0\}.
\]
All inequality signs applied to vectors and matrices are elementwise.
Although in the canonical definition (\ref{eq:entropic}), $a$ and
$b$ need to be probability vectors in the sense that $\sum^{n}_{i=1}a_{i}=\sum^{m}_{j=1}b_{j}=1$,
in the following derivation we can relax this assumption and only
assume that $a>0$, $b>0$, and $\sum^{n}_{i=1}a_{i}=\sum^{m}_{j=1}b_{j}$.
Then we can easily find that $\mathcal{B}_{n}\equiv\Pi(\mathbf{1}_{n},\mathbf{1}_{n})$.

We can then show that the dual problem of (\ref{eq:entropic}) is
\begin{equation}
\max_{\alpha\in\mathbb{R}^{n},\beta\in\mathbb{R}^{m}}\,\mathcal{L}(\alpha,\beta),\quad\mathcal{L}(\alpha,\beta)=-\eta\sum^{n}_{i=1}\sum^{m}_{j=1}\exp\{\eta^{-1}(\alpha_{i}+\beta_{j}-M_{ij})\}+\alpha^{T}a+\beta^{T}b.\label{eq:dual}
\end{equation}
Moreover, if $(\alpha^{*},\beta^{*})$ is an optimal solution to (\ref{eq:dual}),
then the primal solution $T^{*}$ to (\ref{eq:entropic}) can be recovered
as $T^{*}_{ij}=\exp\{(\alpha^{*}_{i}+\beta^{*}_{j}-M_{ij})/\eta\}$.

Given $\beta\in\mathbb{R}^{m}$, let $\alpha^{*}(\beta)$ be the
maximizer of $\mathcal{L}(\alpha,\beta)$ with respect to $\alpha$,
\emph{i.e.}, $\alpha^{*}(\beta)=\arg\max_{\alpha}\:\mathcal{L}(\alpha,\beta)$,
and then we can show that $\alpha^{*}(\beta)$ has a closed-form formula:
\begin{equation}
\alpha^{*}_{i}(\beta)=\eta\log(a_{i})-\eta\log\left[\sum^{m}_{j=1}e^{(\beta_{j}-M_{ij})/\eta}\right],\quad i=1,\ldots,n,\label{eq:alpha_opt}
\end{equation}
where $\alpha^{*}_{i}(\beta)$ is the $i$-th element of the vector
$\alpha^{*}(\beta)$. Similarly, define $\beta^{*}(\alpha)$ to be the maximizer of $\mathcal{L}(\alpha,\beta)$ with respect to $\beta$, $\beta^{*}(\alpha)=\arg\max_{\beta}\:\mathcal{L}(\alpha,\beta)$,
and then we have
\begin{equation}
\beta^{*}_{j}(\alpha)=\eta\log(b_{j})-\eta\log\left[\sum^{n}_{i=1}e^{(\alpha_{i}-M_{ij})/\eta}\right],\quad j=1,\ldots,m.\label{eq:beta_opt}
\end{equation}

With the two partial maximizers $\alpha^{*}(\beta)$ and $\beta^{*}(\alpha)$,
one well-known and natural method to solve (\ref{eq:dual}) is the
block coordinate ascent (BCA) algorithm, which alternately maximizes
one component of the dual variables given the other. Specifically,
given an initial vector $\beta^{(0)}\in\mathbb{R}^{m}$, BCA proceeds
using the following update rule:
\begin{equation}
\alpha^{(k+1)}=\alpha^{*}(\beta^{(k)}),\quad\beta^{(k+1)}=\beta^{*}(\alpha^{(k+1)}),\quad k=0,1,\ldots.\label{eq:bca}
\end{equation}

We then show that the BCA algorithm (\ref{eq:bca}) is equivalent
to the Sinkhorn--Knopp algorithm introduced in (\ref{eq:sinkhorn_knopp}).
Let $u=\exp(\eta^{-1}\alpha)$ and $v=\exp(\eta^{-1}\beta)$, where
the $\exp(\cdot)$ function applies to each element of vectors and
matrices. Clearly, (\ref{eq:alpha_opt}) is equivalent to
\[
u^{*}_{i}(v)=\frac{a_{i}}{\sum^{m}_{j=1}e^{(\beta_{j}-M_{ij})/\eta}}=\frac{a_{i}}{\sum^{m}_{j=1}v_{j}K_{ij}}=\frac{a_{i}}{(Kv)_{i}},
\]
where $K_{ij}=e^{-M_{ij}/\eta}$. Similarly, (\ref{eq:beta_opt})
reduces to
\[
v^{*}_{j}(u)=\frac{b_{j}}{(K^{T}u)_{j}},
\]
and we can compactly write $u^{*}(v)=a\oslash(Kv)$ and $v^{*}(u)=b\oslash(K^{T}u)$,
which exactly recovers the Sinkhorn--Knopp algorithm.

In this sense, we can conclude that projecting a matrix $\exp(R)$
onto the Birkhoff polytope $\mathcal{B}_{n}$ using the Sinkhorn--Knopp
algorithm is equivalent to solving an entropic-regularized OT problem
with cost matrix $M=-R$, regularization parameter $\eta=1$, and
uniform marginal vectors $a=b=\mathbf{1}_{n}$. In other words, if
we can find an efficient algorithm for problem (\ref{eq:dual}), then
the Birkhoff projection problem is automatically solved as a special
case.

\section{Forward Pass: Newton's Method}

\subsection{Second-Order Solver}

At first glance, the dual problem (\ref{eq:dual}) has $(n+m)$ variables,
but we can show that the number of free variables can be reduced to
$(m-1)$. Given a vector $v\in\mathbb{R}^{m}$ and a matrix $A\in\mathbb{R}^{n\times m}$,
denote $v_{-m}=(v_{1},\ldots,v_{m-1})^{T}$, and let $A_{-m}$ be
the matrix after removing the $m$-th column of $A$. We first note
that the dual variables $(\alpha,\beta)$ in (\ref{eq:dual}) have
a redundant degree of freedom, as $\mathcal{L}(\alpha,\beta)\equiv\mathcal{L}(\alpha+c\mathbf{1}_{n},\beta-c\mathbf{1}_{m})$
for any $c\in\mathbb{R}$. Therefore, we can globally set $\beta_{m}=0$,
and always let $\beta=(\beta_{-m},\beta_{m})=(\beta_{-m},0)$. Next,
since $\alpha^{*}(\beta)$ partially maximizes $\mathcal{L}(\alpha,\beta)$
given $\beta$, we can define
\[
f(\beta_{-m})=-\mathcal{L}(\alpha^{*}(\beta),\beta)=\eta\mathbf{1}^{T}_{n}a-[\alpha^{*}(\beta)]^{T}a-\beta^{T}b,
\]
and then maximizing $\mathcal{L}(\alpha,\beta)$ for $(\alpha,\beta)$
is equivalent to minimizing $f(\beta_{-m})$ for $\beta_{-m}$, which
only has $(m-1)$ variables. For brevity, let $x\equiv\beta_{-m}$
be the free variable, and then solving (\ref{eq:dual}) reduces to
\begin{equation}
\min_{x\in\mathbb{R}^{m-1}}\,f(x)\label{eq:min_fx}
\end{equation}
for an $(m-1)$-dimensional variable $x$. The problem (\ref{eq:min_fx})
has some properties favorable for second-order optimization: $f(x)$ is strictly convex and twice-differentiable,
and the minimization problem (\ref{eq:min_fx}) is unconstrained.
This motivates us to consider various optimization techniques for
smooth and unconstrained problems, such as gradient descent and Newton's
method.

To apply these methods, we need to derive the gradient and Hessian
matrix of $f(x)$. In fact, we can prove that
\begin{equation}
\begin{aligned}\nabla f(x) & =[T(\beta)]^{T}_{-m}\mathbf{1}_{n}-b_{-m},\\
\nabla^{2}f(x) & =\eta^{-1}\left\{ \mathbf{diag}([T(\beta)]^{T}_{-m}\mathbf{1}_{n})-[T(\beta)]^{T}_{-m}\mathbf{diag}(a)^{-1}[T(\beta)]_{-m}\right\} ,
\end{aligned}
\label{eq:grad_hess}
\end{equation}
where $T(\beta)$ is a matrix with elements $T_{ij}=\exp\{\eta^{-1}(\alpha^{*}_{i}(\beta)+\beta_{j}-M_{ij})\}$.
Below we show an interesting expression for the $T$ matrix. Given
a matrix $M\in\mathbb{R}^{n\times m}$, let $M_{i\cdot}=(M_{i1},\ldots,M_{im})^{T}$
denote the vector of the $i$-th row of $M$, and $M_{\cdot j}$ be
the $j$-th column of $M$. Define $U\in\mathbb{R}^{n\times m}$ to
be a matrix with entries $U_{ij}=\beta_{j}-M_{ij}$, and then by definition,
\[
T_{ij}=a_{i}\cdot\frac{e^{(\beta_{j}-M_{ij})/\eta}}{\sum^{m}_{k=1}e^{(\beta_{k}-M_{ik})/\eta}}=a_{i}\cdot\frac{e^{U_{ij}/\eta}}{\sum^{m}_{k=1}e^{U_{ik}/\eta}},\quad i=1,\ldots,n,\ j=1,\ldots,m.
\]
This essentially means that
\[
T_{i\cdot}=a_{i}\cdot\mathrm{Softmax}(\eta^{-1}U_{i\cdot}),
\]
where
\[
\mathrm{Softmax}(v)=\left(\frac{v_{1}}{\sum^{m}_{k=1}e^{v_{k}}},\ldots,\frac{v_{m}}{\sum^{m}_{k=1}e^{v_{k}}}\right)^{T}
\]
is the Softmax function for a vector $v=(v_{1},\ldots,v_{m})^{T}$.
Therefore, the $T$ matrix can be obtained by applying the Softmax
function to each row of the matrix $\eta^{-1}U$.

It is well-known that for twice-differentiable convex optimization
problems, Newton's method achieves a fast convergence with a local
quadratic rate. Starting from an initial value $x^{(0)}$, Newton's
method solves (\ref{eq:min_fx}) using the iteration
\[
x^{(k+1)}=x^{(k)}-\gamma_{k}[\nabla^{2}f(x^{(k)})]^{-1}[\nabla f(x^{(k)})],
\]
where $\gamma_{k}$ is the step size at iteration $k$, typically
determined by line search algorithms. Once we have obtained the optimal
point $x^{*}=\beta^{*}_{-m}$, the solution to (\ref{eq:entropic})
is recovered as $T^{*}_{ij}=\exp\{(\alpha^{*}_{i}(\beta^{*})+\beta^{*}_{j}-M_{ij})/\eta\}$,
where $\beta^{*}=(\beta^{*}_{-m},0)$.

For large-scale OT problems, Newton's method is rarely used, since
both the storage and computational cost for computing the Newton direction
$-[\nabla^{2}f(x)]^{-1}[\nabla f(x)]$ is enormous. However, for mHC,
the typical scale of one Birkhoff projection problem is only $4\times4$
or $8\times8$, but the batch size $N$ may be large, which means
that $N$ small and independent Birkhoff projection problems need
to be efficiently solved. In the next two sections, we show that the
$4\times4$ problem is especially suitable for modern GPU hardware.

\subsection{Specialization to $4\times4$ Birkhoff Projection}

\label{subsec:4x4}

As we have shown in Section \ref{subsec:entropic_regularized_ot},
projecting a matrix $\exp(R)$ onto the Birkhoff polytope $\mathcal{B}_{n}$
using the Sinkhorn--Knopp algorithm can be viewed as a special case
of the entropic-regularized OT problem (\ref{eq:entropic}) with $n=m$,
$M=-R$, $\eta=1$, and $a=b=\mathbf{1}_{n}$. In this section, we
consider the case $n=m=4$, which is the setting used by \citet{xie2025mhc}
for mHC implementation. Due to its special structure, the gradient
and Hessian computation can be greatly simplified, and is naturally
fitted to GPU implementation.

In this setting, $R$ is a $4\times4$ matrix, $\alpha=(\alpha_{1},\alpha_{2},\alpha_{3},\alpha_{4})^{T}$
is a $4\times1$ vector, and $x=\beta_{-m}=(\beta_{1},\beta_{2},\beta_{3})^{T}$
is a $3\times1$ vector. We explicitly write
\[
U=\mathbf{1}_{4}\beta^{T}-M=\mathbf{1}_{4}\beta^{T}+R=\begin{bmatrix}\beta_{1}+R_{11} & \beta_{2}+R_{12} & \beta_{3}+R_{13} & \beta_{4}+R_{14}\\
\beta_{1}+R_{21} & \beta_{2}+R_{22} & \beta_{3}+R_{23} & \beta_{4}+R_{24}\\
\beta_{1}+R_{31} & \beta_{2}+R_{32} & \beta_{3}+R_{33} & \beta_{4}+R_{34}\\
\beta_{1}+R_{41} & \beta_{2}+R_{42} & \beta_{3}+R_{43} & \beta_{4}+R_{44}
\end{bmatrix},
\]
and then we have already obtained that
\[
T_{ij}=a_{i}\cdot\frac{e^{U_{ij}/\eta}}{\sum^{m}_{k=1}e^{U_{ik}/\eta}}=\frac{e^{U_{ij}/\eta}}{\sum^{m}_{k=1}e^{U_{ik}/\eta}},\quad i=1,\ldots,n,\ j=1,\ldots,m.
\]
For a practical implementation, we need to avoid the overflow of the
exponentials, so let $\mu_{1}=\max\{U_{11},U_{12},U_{13},U_{14}\},\ldots,\mu_{4}=\max\{U_{41},U_{42},U_{43},U_{44}\}$,
and then we have
\begin{equation}
T_{ij}=\frac{e^{U_{ij}/\eta}}{\sum^{m}_{k=1}e^{U_{ik}/\eta}}=\frac{e^{(U_{ij}-\mu_{i})/\eta}}{\sum^{m}_{k=1}e^{(U_{ik}-\mu_{i})/\eta}},\quad i=1,\ldots,n,\ j=1,\ldots,m.\label{eq:tij_stable}
\end{equation}
Since in (\ref{eq:tij_stable}) each exponential term is upper bounded
by one, computing $T$ using formula (\ref{eq:tij_stable}) is numerically
stable. Then we can compute
\begin{equation}
c=\begin{bmatrix}c_{1}\\
c_{2}\\
c_{3}
\end{bmatrix}=\begin{bmatrix}T_{11}+T_{21}+T_{31}+T_{41}\\
T_{12}+T_{22}+T_{32}+T_{42}\\
T_{13}+T_{23}+T_{33}+T_{43}
\end{bmatrix},\quad g=\begin{bmatrix}g_{1}\\
g_{2}\\
g_{3}
\end{bmatrix}=\begin{bmatrix}c_{1}-b_{1}\\
c_{2}-b_{2}\\
c_{3}-b_{3}
\end{bmatrix}=\begin{bmatrix}c_{1}-1\\
c_{2}-1\\
c_{3}-1
\end{bmatrix},\label{eq:c_g}
\end{equation}
where $g$ is the gradient vector of $f(x)$ at $x=(\beta_{1},\beta_{2},\beta_{3})^{T}$.

To get the expression for $H=\nabla^{2}f(x)$, note that $[T(\beta)]^{T}_{-m}\mathbf{1}_{n}=c$
and $\mathbf{diag}(a)=I_{n}$, so
\[
H=\eta^{-1}\left\{ \mathbf{diag}(c)-[T(\beta)]^{T}_{-m}[T(\beta)]_{-m}\right\} .
\]
Write
\[
T_{-m}=\begin{bmatrix}T_{11} & T_{12} & T_{13}\\
T_{21} & T_{22} & T_{23}\\
T_{31} & T_{32} & T_{33}\\
T_{41} & T_{42} & T_{43}
\end{bmatrix}=[l_{1},l_{2},l_{3}],
\]
and then the lower triangular part of $H$ is
\[
H=\begin{bmatrix}h_{11} & * & *\\
h_{21} & h_{22} & *\\
h_{31} & h_{32} & h_{33}
\end{bmatrix}=\eta^{-1}\left\{ \begin{bmatrix}c_{1}\\
 & c_{2}\\
 &  & c_{3}
\end{bmatrix}-\begin{bmatrix}p_{11} & * & *\\
p_{21} & p_{22} & *\\
p_{31} & p_{32} & p_{33}
\end{bmatrix}\right\} ,
\]
where
\begin{align*}
p_{11} & =l^{T}_{1}l_{1}=T_{11}T_{11}+T_{21}T_{21}+T_{31}T_{31}+T_{41}T_{41},\\
p_{21} & =l^{T}_{2}l_{1}=T_{12}T_{11}+T_{22}T_{21}+T_{32}T_{31}+T_{42}T_{41},\\
p_{31} & =l^{T}_{3}l_{1}=T_{13}T_{11}+T_{23}T_{21}+T_{33}T_{31}+T_{43}T_{41},\\
p_{22} & =l^{T}_{2}l_{2}=T_{12}T_{12}+T_{22}T_{22}+T_{32}T_{32}+T_{42}T_{42},\\
p_{32} & =l^{T}_{3}l_{2}=T_{13}T_{12}+T_{23}T_{22}+T_{33}T_{32}+T_{43}T_{42},\\
p_{33} & =l^{T}_{3}l_{3}=T_{13}T_{13}+T_{23}T_{23}+T_{33}T_{33}+T_{43}T_{43}.
\end{align*}
For $d=(d_{1},d_{2},d_{3})^{T}=-H^{-1}g$, we have the closed-form
expressions:
\begin{align}
d_{1} & =-[(h_{22}h_{33}-h^{2}_{23})g_{1}+(h_{13}h_{23}-h_{12}h_{33})g_{2}+(h_{12}h_{23}-h_{22}h_{13})g_{3}]/\det(H),\nonumber \\
d_{2} & =-[(h_{13}h_{23}-h_{12}h_{33})g_{1}+(h_{11}h_{33}-h^{2}_{13})g_{2}+(h_{12}h_{13}-h_{11}h_{23})g_{3}]/\det(H),\label{eq:newton_direction}\\
d_{3} & =-[(h_{12}h_{23}-h_{22}h_{13})g_{1}+(h_{12}h_{13}-h_{11}h_{23})g_{2}+(h_{11}h_{22}-h^{2}_{12})g_{3}]/\det(H),\nonumber 
\end{align}
where
\[
\det(H)=h_{11}(h_{22}h_{33}-h^{2}_{23})-h_{12}(h_{12}h_{33}-h_{23}h_{13})+h_{13}(h_{12}h_{23}-h_{22}h_{13}).
\]

\subsection{CUDA Implementation}

\label{subsec:cuda}

The various numerical operations introduced in Section \ref{subsec:4x4}
can be efficiently implemented on modern GPU hardware. In this article,
we consider the implementation on the CUDA platform. In its programming
model, every 32 GPU threads form a warp, which is the basic unit for
instruction execution. Therefore, we can fit two $4\times4$ matrices
into one warp, and design parallel algorithms to efficiently compute
linear algebra results. Below we show an example on how to compute
the $c$ vectors from two $4\times4$ matrices using only four basic
GPU instructions.

\begin{figure}[h]
\begin{centering}
\includegraphics[width=0.99\textwidth]{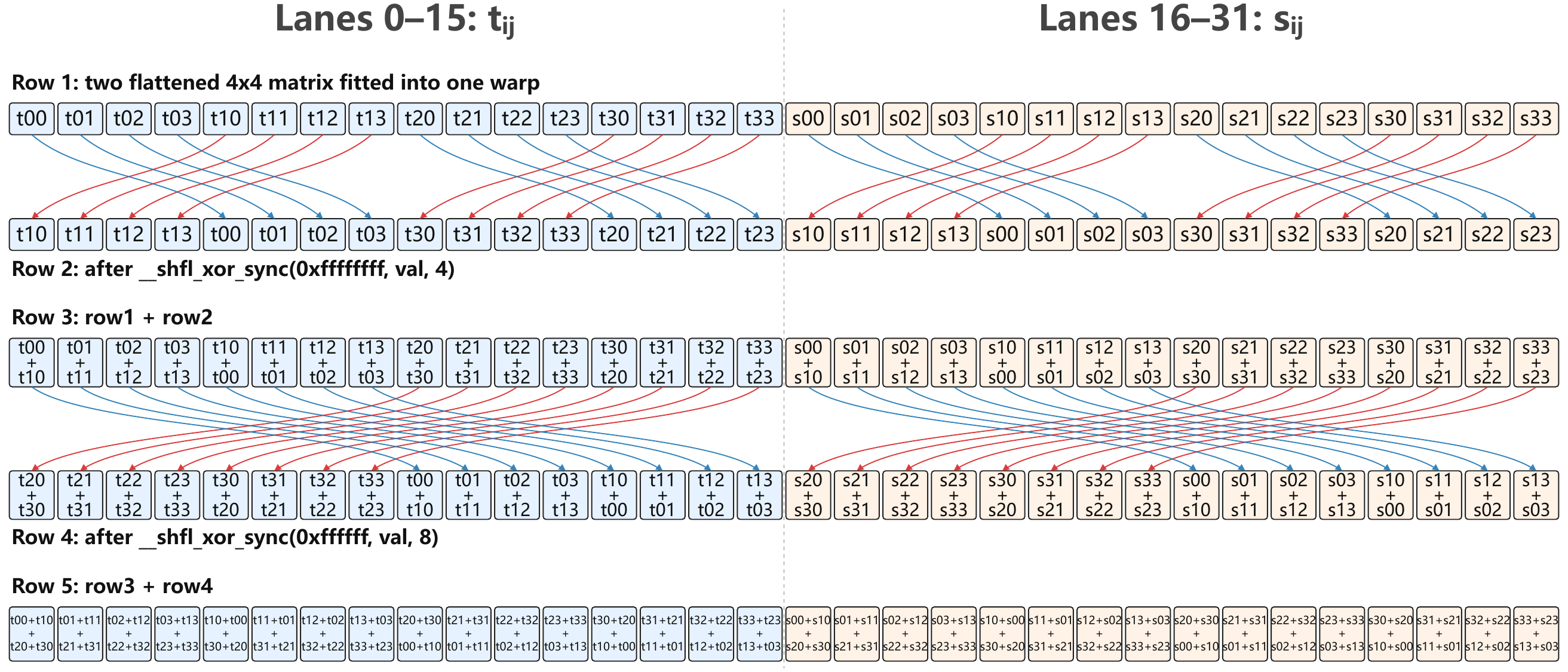}
\par\end{centering}
\caption{Diagram of column sum computation using intra-warp shuffling instructions.}\label{fig:diagram_shuffle}
\end{figure}

Suppose that we label the threads in a warp using indices $0,\ldots,31$,
and we call each of these indices a lane ID. Then we can use lanes
0-15 to process one $4\times4$ matrix, and use lanes 16-31 to process
another, with all 32 threads working simultaneously. Assume that at
some time point of the program, each thread in lanes 0-15 contains
a variable \texttt{val} that holds one element of the $T$ matrix
(\ref{eq:tij_stable}) computed from an $M$ matrix. For example,
in lane 0 \texttt{val} refers to $t_{00}\equiv T_{11}$, in lane 10
\texttt{val} refers to $t_{22}\equiv T_{33}$, etc., where we use
zero-based indices here to align with the convention of the C++ programming
language. Similarly, we assume that lanes 16-31 contain the $T$ values
computed from another $M$ matrix, denoted by $s_{ij}$ to distinguish
from $t_{ij}$. This layout can be visualized by the first row of
Figure \ref{fig:diagram_shuffle}.

We note that each of the $c$ values as in (\ref{eq:c_g}) is the
sum of four $t_{ij}$ or $s_{ij}$ values in the same column, and
we need to compute six $c$ values from two $T$ matrices. To compute
the column sums of $T$, each thread holding $t_{ij}$ must know the
value of \texttt{val} held by other threads in the same matrix column,
which undoubtedly requires the communication among threads. In the
general setting, this can be achieved by the reading and writing of
global memory or shared memory, but it comes with an I/O cost. Fortunately,
for threads in the same warp, there is a register-level shuffling
operation \texttt{\_\_shfl\_xor\_sync()} that allows each thread to read
the value held by another thread at a specific location. The second
row of Figure \ref{fig:diagram_shuffle} illustrates this process:
on each thread containing $t_{ij}$, \texttt{\_\_shfl\_xor\_sync(0xffffffff,
val, 4)} returns the value $t_{i'j}$ from another matrix row $i'$
but with the same column index $j$, and $i'$ is determined by the
shuffling pattern visualized by the red and blue arrows. Then on each
thread, we add these two values together, and do another round of
shuffling using the instruction \texttt{\_\_shfl\_xor\_sync(0xffffffff,
val, 8)}, as illustrated by row 3 and row 4 of Figure \ref{fig:diagram_shuffle}.
Finally, with another adding operation, each thread that originally
contains $t_{ij}$ will obtain $t_{0j}+t_{1j}+t_{2j}+t_{3j}$ at the
end, which is exactly the $j$-th column sum of $T$ and also the
value of $c_{j}$.

We also observe that the two shuffling operations do not disrupt the
isolation between lanes 0-15 and lanes 16-31, which means that all
threads in this warp can compute their own column sums independently
and simultaneously. Since each instruction in this process, either
adding or shuffling, is executed by all 32 threads in the warp simultaneously,
we eventually finish computing the $c$ vectors from two $T$ matrices
using only four basic instructions without any memory I/O.

Using a similar technique, we implement other parts of the algorithm
using only register-level instructions. The complete programming code
can be found in the GitHub repository: \url{https://github.com/yixuan/mHC-proj}.

\section{Backward Pass: Implicit Differentiation}

\subsection{General Case}

In the forward process, we solve the optimization problem (\ref{eq:entropic})
to compute the solution $T^{*}$ given $M$, so $T^{*}=T^{*}(M)$ is
implicitly a function of $M$. Then in the backward process, we need
to compute the derivative of $T^{*}$ with respect to $M$. Suppose
that we have access to the upstream derivative $G$ for $T^{*}$,
\emph{i.e.},
\[
\frac{\partial\ell}{\partial\mathbf{vec}(T^{*})^{T}}=\mathbf{vec}(G)^{T}\in\mathbb{R}^{1\times(nm)}
\]
for some loss function $\ell$, and then by the chain rule, we have
\[
\frac{\partial\ell}{\partial\mathbf{vec}(M)^{T}}=\frac{\partial\ell}{\partial\mathbf{vec}(T^{*})^{T}}\left[\frac{\partial\mathbf{vec}(T^{*})^{T}}{\partial\mathbf{vec}(M)^{T}}+\frac{\partial\mathbf{vec}(T^{*})^{T}}{\partial(x^{*})^{T}}\,\frac{\partial(x^{*})^{T}}{\partial\mathbf{vec}(M)^{T}}\right],
\]
where $x^{*}$ is the optimal solution to (\ref{eq:min_fx}). Our
target in the backward process is to compute $\partial\ell/\partial\mathbf{vec}(M)^{T}$
given $G$ and $T^{*}$, and the key part in deriving the gradient
is the term $\partial(x^{*})^{T}/\partial\mathbf{vec}(M)^{T}$. We
solve this problem using the implicit function theorem (see, for example, Theorem 1B.1 of \citealp{dontchev2009implicit}), as $x^{*}$
relies on $M$ via the optimality condition $\nabla f(x^{*})=\mathbf{0}_{m-1}$,
which can be viewed as an equation for $x^{*}$ and $M$ via some
function $g$, $g(x^{*},M)=\mathbf{0}_{m-1}$. Then by the implicit
function theorem, we have
\[
\frac{\partial(x^{*})^{T}}{\partial\mathbf{vec}(M)^{T}}=-\left[\left.\frac{\partial g(x,M)}{\partial x^{T}}\right|_{x=x^{*}}\right]^{-1}\left[\left.\frac{\partial g(x,M)}{\partial\mathbf{vec}(M)^{T}}\right|_{x=x^{*}}\right].
\]
Finally, we can show that
\[
\frac{\partial\ell}{\partial\mathbf{vec}(M)^{T}}=\mathbf{vec}(D)^{T}\in\mathbb{R}^{1\times(nm)},
\]
where $D$ is an $n\times m$ matrix with the following expression:
\begin{equation}
\begin{aligned}D & =\eta^{-1}\left[\mathbf{diag}((\mu_{r}-T^{*}_{-m}w_{-m})\oslash a)T^{*}+T^{*}\mathbf{diag}(w)-(G\circ T^{*})\right],\\
w & =(w_{-m},0),\quad w_{-m}=\Delta^{-1}\left[\mu_{c}-(T^{*}_{-m})^{T}(\mu_{r}\oslash a)\right],\\
\mu_{r} & =(G\circ T^{*})\mathbf{1}_{m},\quad\mu_{c}=(G\circ T^{*})^{T}_{-m}\mathbf{1}_{n},\\
\Delta & =\mathbf{diag}(b_{-m})-(T^{*})^{T}_{-m}\mathbf{diag}(a)^{-1}(T^{*})_{-m},
\end{aligned}
\label{eq:derivative}
\end{equation}
and $\circ$ stands for the Hadamard product between matrices.

\subsection{Specialization to $4\times4$ Problem}

For the $4\times4$ KL-based Birkhoff projection problem, we can get
the simplified expressions for the derivative. Recall that in the
forward pass, we need to compute the Hessian matrix at every Newton
iteration:
\[
H=\nabla^{2}f(x)=\eta^{-1}\left\{ \mathbf{diag}(c)-[T(\beta)]^{T}_{-m}[T(\beta)]_{-m}\right\} .
\]
Suppose that Newton's method stops at an optimal solution $x^{*}$,
and then we have $c=\mathbf{1}_{3}$ and $T(\beta)=T^{*}$, where
$T^{*}$ is the output of the forward pass. Then we can find that
at $x^{*}$, the Hessian matrix exactly matches the $\Delta$ matrix
in (\ref{eq:derivative}), which means that we can save the Hessian
matrix at the last iteration in the forward pass, and reuse it in
the backward pass. Another applicable choice is to recompute $\Delta$
purely from $T^{*}$, since in this way we avoid saving the $h_{ij}$
variables, thus reducing the memory footprint.

Given the input $4\times4$ matrices $G=(g_{ij})$ and $T^{*}=(t_{ij})$,
first obtain $\Gamma=(\gamma_{ij})=G\circ T^{*}\in\mathbb{R}^{4\times4}$,
and then we can again use the similar technique introduced in Section
(\ref{subsec:cuda}) to compute its row sum vector $\mu_{r}$ and
column sum vector $\mu_{c}$:
\[
\mu_{r}=\begin{bmatrix}\gamma_{11}+\gamma_{12}+\gamma_{13}+\gamma_{14}\\
\gamma_{21}+\gamma_{22}+\gamma_{23}+\gamma_{24}\\
\gamma_{31}+\gamma_{32}+\gamma_{33}+\gamma_{34}\\
\gamma_{41}+\gamma_{42}+\gamma_{43}+\gamma_{44}
\end{bmatrix},\quad\mu_{c}=\begin{bmatrix}\gamma_{11}+\gamma_{21}+\gamma_{31}+\gamma_{41}\\
\gamma_{12}+\gamma_{22}+\gamma_{32}+\gamma_{42}\\
\gamma_{13}+\gamma_{23}+\gamma_{33}+\gamma_{43}
\end{bmatrix}.
\]
In the CUDA implementation, the $\mu_{r}$ vector can be computed
with shuffling instructions \texttt{\_\_shfl\_xor\_sync(0xffffffff,
val, 1)} and \texttt{\_\_shfl\_xor\_sync(0xffffffff, val, 2)}, combined
with two parallel adding operations.

For the $w$ vector, since $a=\mathbf{1}_{4}$, we have $w_{-m}=(w_{1},w_{2},w_{3})^{T}=\Delta^{-1}\left[\mu_{c}-(T^{*}_{-m})^{T}\mu_{r}\right]$,
and then we use the formula (\ref{eq:newton_direction}) to solve
the $3\times3$ linear system. By setting $w_{4}=0$ and computing
$v=(v_{1},v_{2},v_{3},v_{4})^{T}=\mu_{r}-T^{*}_{-m}w_{-m}=\mu_{r}-T^{*}w$,
we eventually obtain
\[
D=\begin{bmatrix}(v_{1}+w_{1}-g_{11})t_{11} & (v_{1}+w_{2}-g_{12})t_{12} & (v_{1}+w_{3}-g_{13})t_{13} & (v_{1}+w_{4}-g_{14})t_{14}\\
(v_{2}+w_{1}-g_{21})t_{21} & (v_{2}+w_{2}-g_{22})t_{22} & (v_{2}+w_{3}-g_{23})t_{23} & (v_{2}+w_{4}-g_{24})t_{24}\\
(v_{3}+w_{1}-g_{31})t_{31} & (v_{3}+w_{2}-g_{32})t_{32} & (v_{3}+w_{3}-g_{33})t_{33} & (v_{3}+w_{4}-g_{34})t_{34}\\
(v_{4}+w_{1}-g_{41})t_{41} & (v_{4}+w_{2}-g_{42})t_{42} & (v_{4}+w_{3}-g_{43})t_{43} & (v_{4}+w_{4}-g_{44})t_{44}
\end{bmatrix}.
\]
Note that $D$ is the derivative with respect to $M$, and we need
to flip its sign if the derivative for $R=-M$ is requested.

\section{Numerical Experiments}

In this section, we conduct numerical experiments to validate both the accuracy and the computational performance of the proposed Birkhoff projection method.
We have implemented the algorithm in both C++ CUDA and TileLang, which we denote by mHC-proj and mHC-proj-TL, respectively.
Overall, we consider the following seven open-source implementations:
\begin{enumerate}
\item \textbf{Vanilla}: a simple implementation of the Sinkhorn--Knopp
algorithm using pure PyTorch code.
\item \textbf{Triton-Sinkhorn}: a CUDA-fused implementation of the Sinkhorn--Knopp
algorithm backed by OpenAI Triton: \url{https://github.com/LottoLottoLotto/triton-sinkhorn}.
\item \textbf{mHC.cu}: a CUDA implementation of mHC, with specialized optimizations
for $n=4$: \url{https://github.com/AndreSlavescu/mHC.cu}.
\item \textbf{TileLangExamples}: a TileLang implementation of the Sinkhorn--Knopp
algorithm adapted from the TileLang examples, with a backward pass using implicit conjugate gradient: \url{https://github.com/tile-ai/tilelang/tree/main/examples/deepseek_mhc}.
\item \textbf{TileKernels}: a TileLang implementation of the Sinkhorn--Knopp
algorithm adapted from the DeepSeek TileKernels implementation: \url{https://github.com/deepseek-ai/TileKernels}.
\item \textbf{mHC-proj-TL}: a TileLang implementation of the proposed second-order
Birkhoff projection solver: \url{https://github.com/yixuan/mHC-proj/tree/master/benchmark/mhc/tilelang}
\item \textbf{mHC-proj}: the proposed second-order solver: \url{https://github.com/yixuan/mHC-proj}.
\end{enumerate}
We first generate a tensor $\tilde{\mathcal{H}}^{\mathrm{res}}$ of
size $N\times4\times4$, and then compute the KL projection of $\exp(\tilde{\mathcal{H}}^{\mathrm{res}})$
onto the Birkhoff polytope using different solvers, resulting in an
$N\times4\times4$ tensor $\mathcal{H}^{\mathrm{res}}$. The elements
of $\tilde{\mathcal{H}}^{\mathrm{res}}$ are generated using various
statistical distributions to reflect different structures and magnitudes
of the residual mappings in practical model training. For the Sinkhorn--Knopp
algorithm, we follow the hyperparameter setting in \citet{xie2025mhc}
to run 20 Sinkhorn--Knopp iterations. For the proposed second-order
solver, we use a convergence tolerance of $10^{-6}$ and a maximum
number of 20 Newton iterations. All experiments are
benchmarked on an NVIDIA RTX 6000 Ada Generation GPU.

In Tables \ref{tab:accuracy_small} and \ref{tab:accuracy_large},
we demonstrate the accuracy of different projection methods on the
computed $\mathcal{H}^{\mathrm{res}}$ tensor.
For the $i$-th instance in $\mathcal{H}^{\mathrm{res}}$, \emph{i.e.},
$T=\mathcal{H}^{\mathrm{res}}_{i}\in\mathbb{R}^{4\times4}$, let $r=T\mathbf{1}_{4}\in\mathbb{R}^{4}$
and $c=T^{T}\mathbf{1}_{4}\in\mathbb{R}^{4}$ be its row sum vector
and column sum vector, respectively. Then we define the marginal error
of $T$ as
\[
\mathrm{Err}(T)=\Vert r-\mathbf{1}_{4}\Vert_{1}+\Vert c-\mathbf{1}_{4}\Vert_{1}=|r_{1}-1|+\cdots+|r_{4}-1|+|c_{1}-1|+\cdots+|c_{4}-1|.
\]
With $N=10000$ instances, we summarize the mean, standard deviation,
median, and maximum value of $\mathrm{Err}(\mathcal{H}^{\mathrm{res}}_{i})$
in Tables \ref{tab:accuracy_small} and \ref{tab:accuracy_large}.

\begin{table}[t]
\caption{Accuracy of different projection methods for small-magnitude
inputs.}\label{tab:accuracy_small}

\centering{}%
\begin{tabular}{cc>{\centering}p{0.075\textwidth}>{\centering}p{0.075\textwidth}>{\centering}p{0.075\textwidth}>{\centering}p{0.075\textwidth}c}
\toprule
Entries & Method & Mean & Std. & Median & Max & \tabularnewline
\midrule
\midrule
\multirow{7}{*}[-0.3cm]{$N(0,1)$} & Vanilla & \cellcolor{Salmon}8.336 & \cellcolor{Salmon}20.01 & \cellcolor{Salmon}7.793 & \cellcolor{Salmon}1381 & \multirow{7}{*}[-0.3cm]{($\times10^{-6}$)}\tabularnewline
\cmidrule{2-6}
 & Triton-Sinkhorn & \cellcolor{lime}0.938 & \cellcolor{Salmon}20.16 & \cellcolor{lime}0.328 & \cellcolor{Salmon}1379 & \tabularnewline
\cmidrule{2-6}
 & mHC.cu & \cellcolor{lime}0.867 & \cellcolor{Salmon}20.17 & \cellcolor{lime}\textbf{0.261} & \cellcolor{Salmon}1379 & \tabularnewline
\cmidrule{2-6}
 & TileLangExamples & \cellcolor{Salmon}8.341 & \cellcolor{Salmon}20.01 & \cellcolor{Salmon}7.793 & \cellcolor{Salmon}1381 & \tabularnewline
\cmidrule{2-6}
 & TileKernels & \cellcolor{Salmon}8.342 & \cellcolor{Salmon}20.01 & \cellcolor{Salmon}7.793 & \cellcolor{Salmon}1381 & \tabularnewline
\cmidrule{2-6}
 & mHC-proj-TL (ours) & \cellcolor{lime}\textbf{0.619} & \cellcolor{lime}\textbf{0.385} & \cellcolor{lime}0.484 & \cellcolor{lime}2.533 & \tabularnewline
\cmidrule{2-6}
 & mHC-proj (ours) & \cellcolor{lime}0.654 & \cellcolor{lime}\textbf{0.385} & \cellcolor{lime}0.527 & \cellcolor{lime}\textbf{2.503} & \tabularnewline
\midrule 
\multirow{7}{*}[-0.3cm]{$\mathrm{Unif}(-1,1)$} & Vanilla & \cellcolor{Salmon}7.790 & \cellcolor{lime}0.141 & \cellcolor{Salmon}7.793 & \cellcolor{Salmon}8.330 & \multirow{7}{*}[-0.3cm]{($\times10^{-6}$)}\tabularnewline
\cmidrule{2-6}
 & Triton-Sinkhorn & \cellcolor{lime}0.342 & \cellcolor{lime}0.103 & \cellcolor{lime}0.328 & \cellcolor{lime}0.864 & \tabularnewline
\cmidrule{2-6}
 & mHC.cu & \cellcolor{lime}\textbf{0.264} & \cellcolor{lime}\textbf{0.083} & \cellcolor{lime}\textbf{0.253} & \cellcolor{lime}\textbf{0.745} & \tabularnewline
\cmidrule{2-6}
 & TileLangExamples & \cellcolor{Salmon}7.806 & \cellcolor{lime}0.105 & \cellcolor{Salmon}7.808 & \cellcolor{Salmon}8.196 & \tabularnewline
\cmidrule{2-6}
 & TileKernels & \cellcolor{Salmon}7.805 & \cellcolor{lime}0.103 & \cellcolor{Salmon}7.808 & \cellcolor{Salmon}8.166 & \tabularnewline
\cmidrule{2-6}
 & mHC-proj-TL (ours) & \cellcolor{lime}0.564 & \cellcolor{Goldenrod}0.359 & \cellcolor{lime}0.447 & \cellcolor{Goldenrod}2.265 & \tabularnewline
\cmidrule{2-6}
 & mHC-proj (ours) & \cellcolor{lime}0.593 & \cellcolor{Goldenrod}0.352 & \cellcolor{lime}0.477 & \cellcolor{Goldenrod}2.325 & \tabularnewline
\bottomrule
\end{tabular}
\end{table}

\begin{table}[t]
\caption{Accuracy of different projection methods for large-magnitude
inputs.}\label{tab:accuracy_large}

\centering{}%
\begin{tabular}{cc>{\centering}p{0.075\textwidth}>{\centering}p{0.075\textwidth}>{\centering}p{0.075\textwidth}>{\centering}p{0.075\textwidth}c}
\toprule
Entries & Method & Mean & Std. & Median & Max & \tabularnewline
\midrule
\midrule
\multirow{7}{*}[-0.3cm]{$N(0,10^{2})$} & Vanilla & \cellcolor{Salmon}72.54 & \cellcolor{Salmon}61.12 & \cellcolor{Salmon}65.13 & \cellcolor{Salmon}832.6 & \multirow{7}{*}[-0.3cm]{($\times10^{-3}$)}\tabularnewline
\cmidrule{2-6}
 & Triton-Sinkhorn & \cellcolor{Salmon}72.50 & \cellcolor{Salmon}61.05 & \cellcolor{Salmon}65.08 & \cellcolor{Salmon}832.3 & \tabularnewline
\cmidrule{2-6}
 & mHC.cu & \cellcolor{Salmon}87.92 & \cellcolor{Salmon}184.4 & \cellcolor{Salmon}65.45 & \cellcolor{Salmon}4000 & \tabularnewline
\cmidrule{2-6}
 & TileLangExamples & \cellcolor{Salmon}50.71 & \cellcolor{Salmon}38.34 & \cellcolor{Salmon}49.53 & \cellcolor{lime}207.7 & \tabularnewline
\cmidrule{2-6}
 & TileKernels & \cellcolor{Salmon}50.71 & \cellcolor{Salmon}38.34 & \cellcolor{Salmon}49.53 & \cellcolor{lime}207.7 & \tabularnewline
\cmidrule{2-6}
 & mHC-proj-TL (ours) & \cellcolor{lime}1.595 & \cellcolor{lime}\textbf{4.428} & \cellcolor{lime}\textbf{0.0009} & \cellcolor{lime}\textbf{91.3} & \tabularnewline
\cmidrule{2-6}
 & mHC-proj (ours) & \cellcolor{lime}\textbf{1.594} & \cellcolor{lime}\textbf{4.428} & \cellcolor{lime}\textbf{0.0009} & \cellcolor{lime}\textbf{91.3} & \tabularnewline
\midrule 
\multirow{7}{*}[-0.3cm]{$\mathrm{Unif}(-10,10)$} & Vanilla & \cellcolor{Salmon}40.02 & \cellcolor{Salmon}37.16 & \cellcolor{Salmon}32.93 & \cellcolor{Salmon}234.0 & \multirow{7}{*}[-0.3cm]{($\times10^{-3}$)}\tabularnewline
\cmidrule{2-6}
 & Triton-Sinkhorn & \cellcolor{Salmon}40.02 & \cellcolor{Salmon}37.16 & \cellcolor{Salmon}32.94 & \cellcolor{Salmon}234.0 & \tabularnewline
\cmidrule{2-6}
 & mHC.cu & \cellcolor{Salmon}40.02 & \cellcolor{Salmon}37.16 & \cellcolor{Salmon}32.94 & \cellcolor{Salmon}234.0 & \tabularnewline
\cmidrule{2-6}
 & TileLangExamples & \cellcolor{Salmon}37.48 & \cellcolor{Salmon}35.17 & \cellcolor{Salmon}30.39 & \cellcolor{Salmon}194.3 & \tabularnewline
\cmidrule{2-6}
 & TileKernels & \cellcolor{Salmon}37.48 & \cellcolor{Salmon}35.17 & \cellcolor{Salmon}30.39 & \cellcolor{Salmon}194.3 & \tabularnewline
\cmidrule{2-6}
 & mHC-proj-TL (ours) & \cellcolor{lime}\textbf{0.159} & \cellcolor{lime}\textbf{1.479} & \cellcolor{lime}\textbf{0.0006} & \cellcolor{lime}\textbf{46.57} & \tabularnewline
\cmidrule{2-6}
 & mHC-proj (ours) & \cellcolor{lime}\textbf{0.159} & \cellcolor{lime}\textbf{1.479} & \cellcolor{lime}\textbf{0.0006} & \cellcolor{lime}\textbf{46.57} & \tabularnewline
\bottomrule
\end{tabular}
\end{table}

Table \ref{tab:accuracy_small} shows that when the
entries of $\tilde{\mathcal{H}}^{\mathrm{res}}$
have a relatively small magnitude, for example, $\tilde{\mathcal{H}}^{\mathrm{res}}\sim N(0,1)$
and $\tilde{\mathcal{H}}^{\mathrm{res}}\sim\mathrm{Unif}(-1,1)$,
all methods have reasonably small mean and median error values. However,
in the $N(0,1)$ case, Sinkhorn--Knopp methods demonstrate enormous
worst-case errors, with the maximum value approximately 1000 times
larger than the mean. In contrast, mHC-proj and mHC-proj-TL
show highly consistent errors, whose maximum values are at the same
order as the mean and median.

The advantage of mHC-proj is substantially more evident when the entries
of $\tilde{\mathcal{H}}^{\mathrm{res}}$ have a large magnitude. For
example, in the $N(0,10^{2})$ and $\mathrm{Unif}(-10,10)$ cases as shown in Table \ref{tab:accuracy_large}, the errors of mHC-proj are
several orders of magnitude smaller than the Sinkhorn-based methods,
especially for the median. The TileLang Sinkhorn variants improve
some large-magnitude Sinkhorn errors, but
they remain much less accurate than the proposed solvers.
Meanwhile, mHC-proj-TL closely matches the accuracy of the CUDA mHC-proj
implementation. This finding suggests that in these cases, 20 Sinkhorn--Knopp
iterations may be insufficient to output an accurate projected residual
mapping, whereas in general, the second-order solver converges much
faster and gives well-controlled marginal errors.

\begin{table}[t]
\caption{Median normalized computational time of different projection
methods. Short column labels denote Triton-Sinkhorn (Triton), TileLangExamples (TLE), TileKernels (TK), mHC-proj-TL (Proj-TL),
and mHC-proj (Proj); Fwd. and Fwd.+Bwd. denote the forward pass and
forward--backward computation, respectively.}\label{tab:runtime}

\centering{}%
\setlength{\tabcolsep}{5pt}%
\begin{tabular}{ccccccccc}
\toprule
Feature & Batch & Vanilla & Triton & mHC.cu & TLE & TK & Proj-TL & Proj\tabularnewline
\midrule
\multirow{5}{*}{Fwd.} & $0.5K$ & 49.778 & 9.449 & 2.908 & 3.092 & 2.862 & 2.925 & \textbf{1.000}\tabularnewline
 & $2K$ & 49.974 & 18.811 & 2.977 & 3.106 & 2.867 & 2.946 & \textbf{1.000}\tabularnewline
 & $8K$ & 49.289 & 89.064 & 4.307 & 3.644 & 2.830 & 2.859 & \textbf{1.000}\tabularnewline
 & $32K$ & 22.879 & 197.626 & 2.354 & 5.399 & 2.262 & 1.377 & \textbf{1.000}\tabularnewline
 & $128K$ & 12.419 & 337.584 & 1.674 & 6.887 & 3.001 & 1.235 & \textbf{1.000}\tabularnewline
\midrule
\multirow{5}{*}{Fwd.+Bwd.} & $0.5K$ & 126.380 & 9.754 & 3.548 & 3.066 & 3.058 & 3.071 & \textbf{1.000}\tabularnewline
 & $2K$ & 125.481 & 12.835 & 3.595 & 3.046 & 3.034 & 3.048 & \textbf{1.000}\tabularnewline
 & $8K$ & 124.970 & 56.654 & 4.367 & 3.076 & 8.498 & 3.048 & \textbf{1.000}\tabularnewline
 & $32K$ & 103.678 & 213.621 & 7.809 & 4.984 & 22.028 & 2.497 & \textbf{1.000}\tabularnewline
 & $128K$ & 51.518 & 376.858 & 22.310 & 6.329 & 28.345 & 1.208 & \textbf{1.000}\tabularnewline
\bottomrule
\end{tabular}
\end{table}

Next, we show that the proposed mHC-proj solver not only generates
accurate Birkhoff projections, but also achieves a higher computational
efficiency compared with the Sinkhorn-based implementations. We fix
the matrix entry distribution to be $N(0,10^{2})$, and benchmark
the projection methods with different batch sizes, $N=0.5K,2K,8K,32K,128K$,
where $1K=1024$. In our experiment, each configuration runs repeatedly,
and the median values of the run times across repetitions are reported.
We normalize the results such that the CUDA mHC-proj
implementation always has one unit of run time.

In Table \ref{tab:runtime}, we show the benchmark results for both
the forward pass and the forward--backward computation. It is clear
from the table that all Sinkhorn-based
methods have median normalized run times larger than one, meaning
that mHC-proj is able to achieve a higher output accuracy
using less computing time. The TileLang implementation mHC-proj-TL is slower than the CUDA mHC-proj in these benchmarks,
but it is still competitive with or faster than most Sinkhorn baselines,
especially at larger batch sizes. The gap between mHC-proj and
other implementations is greatly enlarged if the backward pass is
included.

In the case of $128K$ batch size, CUDA mHC-proj exhibits more than
$20\times$ acceleration over the DeepSeek TileKernels implementation, and mHC-proj-TL remains close to the CUDA implementation while
achieving nearly identical accuracy, as has been shown in Tables
\ref{tab:accuracy_small} and \ref{tab:accuracy_large}.

\section{Conclusion}

mHCs restore the stability of HCs by enforcing a doubly stochastic
constraint on the residual mixing matrix, but this benefit hinges
on repeatedly solving a KL-based Birkhoff projection problem at very
high frequency during training and inference. Standard implementations
rely on a fixed-budget Sinkhorn--Knopp solver and typically differentiate
through unrolled iterations, which can introduce non-trivial overhead.
Moreover, on challenging inputs, an insufficient number of Sinkhorn--Knopp
iterations may produce inaccurate projections that undermine the norm
control properties that mHC seeks to guarantee.

In this work, we develop an acceleration framework tailored to the
practically important $4\times4$ setting. By exploiting the small-scale
structure of the entropic-regularized OT formulation, we reformulate
the problem into a three-dimensional unconstrained convex
problem and apply Newton's method with closed-form gradient and
Hessian, achieving fast convergence and high accuracy. For training-time
differentiation, we replace the unrolled backpropagation with implicit
differentiation, yielding an exact and memory-efficient backward pass.
Finally, we translate these algorithmic advantages into a GPU-efficient
realization via a warp-level CUDA kernel that minimizes memory traffic
and kernel-launch overhead.

Empirically, our algorithm produces substantially more reliable doubly
stochastic projections, especially when the entries of $\tilde{\mathcal{H}}^{\mathrm{res}}$
have a large magnitude, and it achieves significant speedups over
representative Sinkhorn-based baselines. These results suggest that
accurate, second-order, and hardware-aware projection solvers can
be a key enabler for scaling mHC-like architectures where stability
is enforced through frequent manifold projections.

\bibliographystyle{apalike}
\bibliography{ref}

\appendix

\section{Mathematical Proofs}

\subsection{Proof of (\ref{eq:birkhoff_projection})}
\begin{proof}
The Lagrangian of the constrained minimization problem in (\ref{eq:birkhoff_projection})
is
\[
F=\sum_{i,j}\left\{ P_{ij}\log\left(\frac{P_{ij}}{A_{ij}}\right)-P_{ij}+A_{ij}\right\} -\sum_{i}\alpha_{i}\left(\sum_{j}P_{ij}-1\right)-\sum_{j}\beta_{j}\left(\sum_{i}P_{ij}-1\right),
\]
where $\alpha=(\alpha_{1},\ldots,\alpha_{n})^{T}$ and $\beta=(\beta_{1},\ldots,\beta_{n})^{T}$
are dual variables. Taking the derivative of $F$ with respect to
$P_{ij}$ and equating it to zero, we have
\[
\log\frac{P_{ij}}{A_{ij}}+1-1-\alpha_{i}-\beta_{j}=0,\quad i,j=1,\ldots,n,
\]
which leads to $P_{ij}=e^{\alpha_{i}}A_{ij}e^{\beta_{j}}$. Clearly,
taking $D_{1}=\mathbf{diag}(\alpha_{1},\ldots,\alpha_{n})$ and $D_{2}=\mathbf{diag}(\beta_{1},\ldots,\beta_{n})$
gives the desired result.
\end{proof}

\subsection{Proof of (\ref{eq:dual})}
\begin{proof}
Introduce two dual variables $\alpha\in\mathbb{R}^{n}$ and $\beta\in\mathbb{R}^{m}$,
and the Lagrangian of (\ref{eq:entropic}) is
\[
F(T,\alpha,\beta)=\langle T,M\rangle-\eta\cdot h(T)-\langle\alpha,T\mathbf{1}_{m}-a\rangle-\langle\beta,T^{T}\mathbf{1}_{n}-b\rangle.
\]
Take the derivative of $F(T,\alpha,\beta)$ with respect to $T_{ij}$,
and we have
\[
\frac{\partial F(T,\alpha,\beta)}{\partial T_{ij}}=M_{ij}+\eta\log(T_{ij})-\alpha_{i}-\beta_{j}.
\]
The first-order optimality condition then yields $T_{ij}=\exp\{(\alpha_{i}+\beta_{j}-M_{ij})/\eta\}$.
Take the expression of $T_{ij}$ back to the Lagrangian, and then
we obtain the dual form (\ref{eq:dual}).
\end{proof}

\subsection{Proof of (\ref{eq:grad_hess})}
\begin{proof}
Clearly,
\begin{align*}
\frac{\partial f}{\partial\beta_{j}} & =-\sum_{i}a_{i}\cdot\frac{\partial\alpha^{*}_{i}(\beta)}{\partial\beta_{j}}-b_{j}=-\sum_{i}a_{i}\cdot\left(-\eta\cdot\frac{\eta^{-1}e^{(\beta_{j}-M_{ij})/\eta}}{\sum^{m}_{k=1}e^{(\beta_{k}-M_{ik})/\eta}}\right)-b_{j}\\
 & =\sum_{i}a_{i}\cdot\left(\frac{e^{(\beta_{j}-M_{ij})/\eta}}{\sum^{m}_{k=1}e^{(\beta_{k}-M_{ik})/\eta}}\right)-b_{j}\\
 & =\sum_{i}\exp\left\{ \log(a_{i})-\log\left(\sum^{m}_{k=1}e^{(\beta_{k}-M_{ik})/\eta}\right)+(\beta_{j}-M_{ij})/\eta\right\} -b_{j}\\
 & =\sum_{i}\exp\left\{ (\alpha^{*}_{i}(\beta)+\beta_{j}-M_{ij})/\eta\right\} -b_{j}=\sum_{i}T_{ij}-b_{j},
\end{align*}
which gives $\nabla f(x)=[T(\beta)]^{T}_{-m}\mathbf{1}_{n}-b_{-m}$.
Furthermore,
\begin{align*}
\frac{\partial^{2}f}{\partial\beta_{j}\partial\beta_{j}} & =\eta^{-1}\sum_{i}T_{ij}\cdot\left(\frac{\partial\alpha^{*}_{i}(\beta)}{\partial\beta_{j}}+1\right)=\eta^{-1}\sum_{i}T_{ij}\cdot\left(-\frac{e^{(\beta_{j}-M_{ij})/\eta}}{\sum^{m}_{k=1}e^{(\beta_{k}-M_{ik})/\eta}}+1\right)\\
 & =\eta^{-1}\sum_{i}T_{ij}\cdot\left(-T_{ij}/a_{i}+1\right)=\eta^{-1}\left(\sum_{i}T_{ij}-\sum_{i}T^{2}_{ij}/a_{i}\right),\\
\frac{\partial^{2}f}{\partial\beta_{j}\partial\beta_{k}} & =\eta^{-1}\sum_{i}T_{ij}\cdot\frac{\partial\alpha^{*}_{i}(\beta)}{\partial\beta_{k}}=-\eta^{-1}\sum_{i}T_{ij}T_{ik}/a_{i},\quad k\neq j,
\end{align*}
and then we obtain $\nabla^{2}f(x)=\eta^{-1}\left\{ \mathbf{diag}([T(\beta)]^{T}_{-m}\mathbf{1}_{n})-[T(\beta)]^{T}_{-m}\mathbf{diag}(a)^{-1}[T(\beta)]_{-m}\right\} $.
\end{proof}

\subsection{Proof of \eqref{eq:derivative}}

Let $D_{1},D_{2}\in\mathbb{R}^{n\times m}$ denote two matrices such
that
\begin{align*}
\mathbf{vec}(D_{1})^{T} & =\frac{\partial\ell}{\partial\mathbf{vec}(T^{*})^{T}}\:\frac{\partial\mathbf{vec}(T^{*})^{T}}{\partial\mathbf{vec}(M)^{T}}=\mathbf{vec}(G)^{T}\left[\frac{\partial\mathbf{vec}(T^{*})^{T}}{\partial\mathbf{vec}(M)^{T}}\right],\\
\mathbf{vec}(D_{2})^{T} & =\mathbf{vec}(G)^{T}\left[\frac{\partial\mathbf{vec}(T^{*})^{T}}{\partial(x^{*})^{T}}\,\frac{\partial(x^{*})^{T}}{\partial\mathbf{vec}(M)^{T}}\right],
\end{align*}
and then
\[
\frac{\partial\ell}{\partial\mathbf{vec}(M)^{T}}=\mathbf{vec}(D_{1})^{T}+\mathbf{vec}(D_{2})^{T}=\mathbf{vec}(D_{1}+D_{2})^{T},
\]

First note that $T^{*}_{ij}=\exp\{(\alpha^{*}_{i}(\beta^{*},M)+\beta^{*}_{j}-M_{ij})/\eta\}$,
where we slightly modify the definition for $\alpha^{*}_{i}(\cdot)$
to emphasize that it depends on both $\beta^{*}$ and $M$:
\[
\alpha^{*}_{i}(\beta,M)=\eta\log(a_{i})-\eta\log\left[\sum^{m}_{k=1}e^{(\beta_{k}-M_{ik})/\eta}\right].
\]
Then we can show that
\begin{align*}
\frac{\partial\alpha^{*}_{i}}{\partial\beta^{*}_{j}} & =-\eta\cdot\frac{\eta^{-1}\cdot e^{(\beta_{j}-M_{ij})/\eta}}{\sum^{m}_{k=1}e^{(\beta_{k}-M_{ik})/\eta}}=-\frac{e^{(\beta_{j}-M_{ij})/\eta}}{\sum^{m}_{k=1}e^{(\beta_{k}-M_{ik})/\eta}}=-T^{*}_{ij}/a_{i},\\
\frac{\partial\alpha^{*}_{i}}{\partial M_{ij}} & =-\eta\cdot\frac{-\eta^{-1}\cdot e^{(\beta_{j}-M_{ij})/\eta}}{\sum^{m}_{k=1}e^{(\beta_{k}-M_{ik})/\eta}}=T^{*}_{ij}/a_{i},\\
\frac{\partial\alpha^{*}_{i}}{\partial M_{sk}} & =0,\quad s\neq i.
\end{align*}
As a result,
\[
\frac{\partial T^{*}_{sk}}{\partial M_{ij}}=\eta^{-1}T^{*}_{sk}\left(\frac{\partial\alpha_{s}}{\partial M_{ij}}-\frac{\partial M_{sk}}{\partial M_{ij}}\right)=\begin{cases}
\eta^{-1}T^{*}_{ij}(T^{*}_{ij}/a_{i}-1), & s=i,k=j,\\
\eta^{-1}T^{*}_{ik}T^{*}_{ij}/a_{i}, & s=i,k\neq j,\\
0, & s\neq i,
\end{cases}
\]
and then
\begin{align*}
(D_{1})_{ij} & =\sum_{s,k}G_{sk}\cdot\frac{\partial T^{*}_{sk}}{\partial M_{ij}}=\sum_{k}G_{ik}\cdot\frac{\partial T^{*}_{ik}}{\partial M_{ij}}\\
 & =\eta^{-1}T^{*}_{ij}/a_{i}\sum_{k}G_{ik}T^{*}_{ik}-\eta^{-1}G_{ij}T^{*}_{ij}.
\end{align*}
In matrix form, we have
\[
D_{1}=\eta^{-1}\mathbf{diag}(\mu_{r}\oslash a)T^{*}-\eta^{-1}(G\circ T^{*}),\quad\mu_{r}=(G\circ T^{*})\mathbf{1}_{m}.
\]

On the other hand, since $T^{*}_{ij}=\exp\{(\alpha^{*}_{i}(\beta^{*},M)+\beta^{*}_{j}-M_{ij})/\eta\}$
and $\partial\alpha^{*}_{i}/\partial\beta^{*}_{j}=-T^{*}_{ij}/a_{i}$,
we have
\begin{align*}
\frac{\partial T^{*}_{ij}}{\partial\beta^{*}_{j}} & =\eta^{-1}T^{*}_{ij}\left(\frac{\partial\alpha^{*}_{i}}{\partial\beta^{*}_{j}}+1\right)=\eta^{-1}T^{*}_{ij}(1-T^{*}_{ij}/a_{i}),\quad j=1,\ldots,m-1,\\
\frac{\partial T^{*}_{ij}}{\partial\beta^{*}_{k}} & =\eta^{-1}T^{*}_{ij}\cdot\frac{\partial\alpha^{*}_{i}}{\partial\beta^{*}_{k}}=-\eta^{-1}T^{*}_{ij}T^{*}_{ik}/a_{i},\quad k=1,\ldots,m-1,k\neq j.
\end{align*}
Therefore, let
\[
v^{T}=\frac{\partial\ell}{\partial\mathbf{vec}(T^{*})^{T}}\,\frac{\partial\mathbf{vec}(T^{*})^{T}}{\partial(x^{*})^{T}}=\mathbf{vec}(G)^{T}\left[\frac{\partial\mathbf{vec}(T^{*})^{T}}{\partial(x^{*})^{T}}\right],
\]
and then
\[
v_{k}=\sum_{i,j}G_{ij}\cdot\frac{\partial T^{*}_{ij}}{\partial\beta^{*}_{k}}=\eta^{-1}\sum_{i}G_{ik}T^{*}_{ik}-\eta^{-1}\sum_{i,j}G_{ij}T^{*}_{ij}T^{*}_{ik}/a_{i},\quad k=1,\ldots,m-1.
\]
In matrix form, we have
\begin{align*}
v & =\eta^{-1}\mu_{c}-\eta^{-1}(T^{*}_{-m})^{T}(\mu_{r}\oslash a),\\
\mu_{c} & =(G\circ T^{*})^{T}_{-m}\mathbf{1}_{n}.
\end{align*}

Now consider $\partial(x^{*})^{T}/\partial\mathbf{vec}(M)^{T}$. Since
$x^{*}$ is the solution to (\ref{eq:min_fx}), it satisfies the optimality
condition $\nabla f(x^{*})=\mathbf{0}_{m-1}$, which expands to
\[
g(x^{*},M)\coloneqq T^{*}_{-m}\mathbf{1}_{n}-b_{-m}=\mathbf{0}_{m-1},
\]
where
\begin{align*}
T^{*}_{ij} & =\exp\{(\alpha^{*}_{i}(\beta^{*},M)+\beta^{*}_{j}-M_{ij})/\eta\},\\
x^{*}_{j} & =\beta^{*}_{j},\quad j=1,\ldots,m-1.
\end{align*}
By the implicit function theorem,
\[
\frac{\partial(x^{*})^{T}}{\partial\mathbf{vec}(M)^{T}}=-\left[\left.\frac{\partial g(x,M)}{\partial x^{T}}\right|_{x=x^{*}}\right]^{-1}\left[\left.\frac{\partial g(x,M)}{\partial\mathbf{vec}(M)^{T}}\right|_{x=x^{*}}\right]\coloneqq-g^{-1}_{x}g_{M}.
\]
We have already obtained that
\[
g_{x}=\nabla^{2}f(x^{*})=\eta^{-1}\left\{ \mathbf{diag}((T^{*})^{T}_{-m}\mathbf{1}_{n})-(T^{*})^{T}_{-m}\mathbf{diag}(a)^{-1}(T^{*})_{-m}\right\} \coloneqq\eta^{-1}\Delta,
\]
and observe that $(T^{*})^{T}_{-m}\mathbf{1}_{n}=b_{-m}$, so
\[
\mathbf{vec}(D_{2})^{T}=-v^{T}g^{-1}_{x}g_{M}=-\left[\mu_{c}-(T^{*}_{-m})^{T}(\mu_{r}\oslash a)\right]^{T}\Delta^{-1}g_{M}\coloneqq-w^{T}_{-m}g_{M},
\]
and we define $w_{m}=0$.

Recall that $\partial\alpha^{*}_{i}/\partial M_{ij}=T^{*}_{ij}/a_{i}$,
and $\partial\alpha^{*}_{s}/\partial M_{ij}=0$ for $s\neq i$. Then
we have
\begin{align*}
\frac{\partial g_{k}}{\partial M_{ij}} & =\frac{\partial\left(\sum_{s}T^{*}_{sk}-b_{k}\right)}{\partial M_{ij}}=\sum_{s}\frac{\partial T^{*}_{sk}}{\partial M_{ij}}=\eta^{-1}\sum_{s}T^{*}_{sk}\left(\frac{\partial\alpha^{*}_{s}}{\partial M_{ij}}-\frac{\partial M_{sk}}{\partial M_{ij}}\right)\\
 & =\eta^{-1}T^{*}_{ik}\left(\frac{\partial\alpha^{*}_{i}}{\partial M_{ij}}-\frac{\partial M_{ik}}{\partial M_{ij}}\right)=\begin{cases}
\eta^{-1}T^{*}_{ij}(T^{*}_{ij}/a_{i}-1), & k=j,\\
\eta^{-1}T^{*}_{ik}T^{*}_{ij}/a_{i}, & k\neq j.
\end{cases}
\end{align*}
As a result,
\[
(D_{2})_{ij}=-\sum^{m-1}_{k=1}w_{k}\cdot\frac{\partial g_{k}}{\partial M_{ij}}=\begin{cases}
\eta^{-1}w_{j}T^{*}_{ij}-\eta^{-1}T^{*}_{ij}/a_{i}\sum^{m-1}_{k=1}w_{k}T^{*}_{ik}, & j\neq m,\\
-\eta^{-1}T^{*}_{ij}/a_{i}\sum^{m-1}_{k=1}w_{k}T^{*}_{ik}, & j=m.
\end{cases}
\]
Since we have defined $w_{m}=0$, we can uniformly write
\[
(D_{2})_{ij}=\eta^{-1}w_{j}T^{*}_{ij}-\eta^{-1}T^{*}_{ij}/a_{i}\sum^{m-1}_{k=1}w_{k}T^{*}_{ik}.
\]
In matrix form, it is expressed as
\[
D_{2}=\eta^{-1}T^{*}\mathbf{diag}(w)-\eta^{-1}\mathbf{diag}((T^{*}_{-m}w_{-m})\oslash a)T^{*}.
\]
Overall, we can show that
\begin{align*}
\eta(D_{1}+D_{2}) & =\mathbf{diag}(\mu_{r}\oslash a)T^{*}-(G\circ T^{*})+T^{*}\mathbf{diag}(w)-\mathbf{diag}((T^{*}_{-m}w_{-m})\oslash a)T^{*}\\
 & =\mathbf{diag}((\mu_{r}-T^{*}_{-m}w_{-m})\oslash a)T^{*}+T^{*}\mathbf{diag}(w)-(G\circ T^{*}),
\end{align*}
which gives the desired result.
\end{document}